\documentclass[a4paper,11pt]{article}
\usepackage{aaskaiid}
\usepackage{orcidlink}
\usepackage{graphicx}
\usepackage[caption=false]{subfig}
\usepackage[T1]{fontenc}
\usepackage{ragged2e}
\usepackage{natbib}
\usepackage{amsmath}
\usepackage{float}

\title{Probing magnetized accretion disk-jet systems: stellar mass to supermassive black holes}
\ShortTitle{Probing magnetized accretion disk-jet systems}
\author[1]{Mayank Pathak\orcidlink{0000-0002-4834-4211}}
\ShortName{Pathak, Raha \& Mukhopadhyay} % shortened name list for header 
\author[2]{Rohan Raha\orcidlink{0009-0002-2354-2884}}
\author[1,2]{Banibrata Mukhopadhyay\orcidlink{0000-0002-3020-9513}}
\affiliation[1]{Joint Astronomy Programme, Department of Physics, Indian Institute of Science, Bengaluru 560012, India}
\emailAdd{mayankpathak@iisc.ac.in}
\affiliation[2]{Department of Physics, Indian Institute of Science, Bengaluru 560012, India}
\emailAdd{raharohan@iisc.ac.in}
\emailAdd{bm@iisc.ac.in}

\abstract{Ubiquitous nature of accretion disks and associated jets in modern astrophysics is extreme for black holes. The current state-of-the-art of black hole activities lies with modeling of underlying general relativistic magnetohydrodynamic (GRMHD) flows. These simulations have shown the importance of magnetic fields in the generation of outflows/jets and the overall dynamical evolution of the accretion flow. They also reveal critical insights into mechanisms that influence accretion dynamics, jet formation and stability. This further sheds light on the underlying magnetic field configurations based on magnetic field saturation leading to Standard and Normal Evolution: SANE, and Magnetically Arrested Disk: MAD. By employing SKA's high-resolution imaging and sensitivity, we can directly compare simulation outcomes with observational data, validating our models and enhancing our understanding of these phenomena. Key to this investigation is the examination of magnetic fields and their associated polarization signatures. Comparing the observational data from SKA with GRMHD simulations will facilitate a deeper analysis of the polarization properties, which can reveal the magnetic field geometry and dynamics in these extreme environments. The VLBI capabilities of SKA will prove instrumental in understanding jet morphologies and spectra of these systems due to its high spatial resolution. Collating these observations with GRMHD simulations will lead to a better understanding of the jet generation mechanisms and their interaction with ambient medium. By integrating advanced GRMHD simulations with SKA's capabilities, we aim to bridge theoretical predictions and observations, ultimately contributing to a more comprehensive understanding of the behavior of accreting black holes and their jets.}
\begin{document}
\maketitle
\section{Introduction}
Accretion onto gravitational objects leads to some of the most enigmatic and dynamically rich systems. The plethora of hydrodynamic and magnetic effects present in these systems have led to several observational features. Most of these effects, however, occur at very small scales, particularly for compact objects, where general relativity (GR) becomes important and thus these systems cannot be probed by conventional Newtonian mechanics. To tackle this issue, several general relativistic magnetohydrodynamic (GRMHD) codes \citep{harm, echo, athena, bhac}, have been developed and have been used to simulate accretion systems which help study their GR based properties.

With the advent of the event horizon telescope (EHT) observations \citep{eht1}, the importance of very long baseline interferometry (VLBI) and GRMHD simulations became even more evident. Suites of GRMHD simulations with varying parameters, like black hole (BH) spin, magnetic field topology etc., have been used to mimic the EHT observations by generating synthetic photon rings. These studies have been used to constrain various physical properties of the observed BH accretion system.

BH accretion systems are often characterized based on the mass of the central BH, e.g., BH X-ray binaries (for stellar mass BHs) and active galactic nuclei (AGNs) (for supermassive BHs). X-ray binaries comprise of a main sequence companion star along with a stellar mass BH. The accretion onto this BH is driven by Roche lobe overflow from the companion star, resulting in the formation of an accretion disk. In AGNs, however, matter from the surrounding medium accretes onto the supermassive BH at the galactic center. Both these systems, though different in construction and scale, exhibit radio jets.  

Thermal emission from the accretion disk of these systems has been well modeled by multi-colour blackbody spectra \citep{ss73, nt73}. 
Jets, however, primarily emit non-thermal radiation which can have several generation mechanisms like synchrotron, inverse-Comptonization, shock heating etc \citep{falcke}. In fact, the jet emitting accretion flow also exhibits nonthermal optically thin emissions.  

The formation of jets in accretion systems has been well explained by the Blandford-Znajek (BZ) \citep{bz} and the Blandford-Payne (BP) \citep{bp} mechanisms. The former argues for the effect of the BH ergosphere to be the launching mechanism behind jet formation, while the latter explains the jet launching being powered by the angular momentum reservoir of the disk. Both mechanisms, however, rely on poloidal magnetic fields to support the jet flow and on toroidal magnetic fields for the eventual collimation of the jet.

With the SKA's high resolution, sensitivity and polarization measuring capabilities, it will be possible to probe various jet properties in detail, very close to the central BH. The properties of sources obtained via these observations can be used to model the GRMHD setup. The accretion and jet details close to the BH can then be calculated from the simulation results. Hence, comparing the observations with GRMHD simulation will reveal information about various system properties. 

\section{GRMHD simulations}\label{grmhd}
We consider the Black Hole Accretion Code (BHAC) \citep{bhac} for 2D and H-AMR \citep{Liska_2022} for 3D simulations. Due to the inherently GR nature of accretion flows near BHs, MHD equations have to be solved in the framework of general relativity. GRMHD simulations start with a central BH surrounded by matter density, usually in the form of a torus. The size of the torus and in turn the amount of matter in the domain can be controlled such that the system does not run out of matter during the course of the simulation run. The code does not evolve the background spacetime geometry itself due to the low density of the flow and the reasonably short time evolution during which substantial change in the spacetime parameters is not expected. 

The equations solved by the code are:
\begin{equation}
\begin{aligned}
    \nabla_\mu(\rho u^\mu)&=0,\\
    \nabla_\mu T^{\mu\nu}&=0,\\
    \nabla_{\mu}\hspace{0.01in}^*F^{\mu\nu}&=0,   
\end{aligned}
\end{equation}
where $u^{\mu}$ is the four-velocity, $T^{\mu\nu}$ is the stress-energy tensor and $^{*}F^{\mu\nu}$ is the dual Faraday tensor. Here, $\mu$ and $\nu$ are spacetime indices such as $t,r,\theta,\phi$.
\begin{figure*}
\centering
\subfloat[MAD]{
\includegraphics[width=\textwidth]{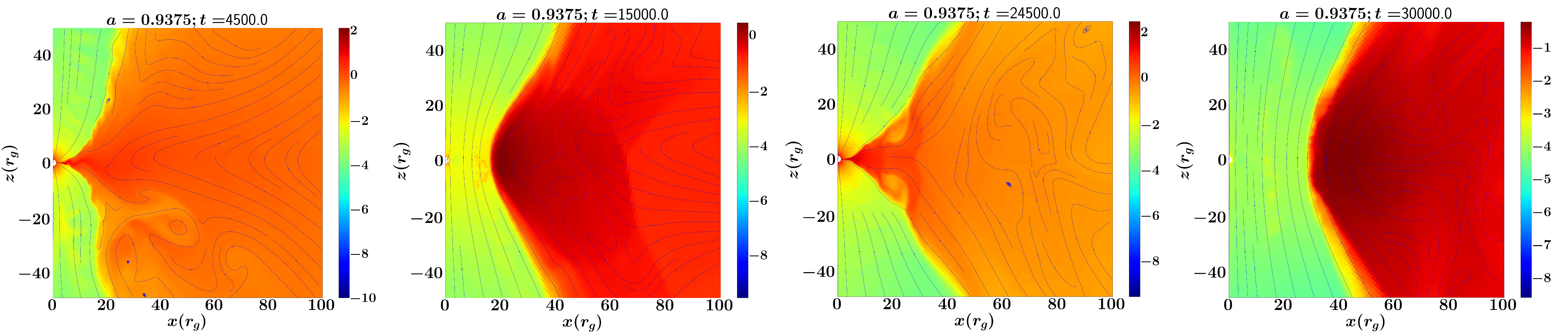}\label{mad}
}

\subfloat[SANE]{
\includegraphics[width=\textwidth]{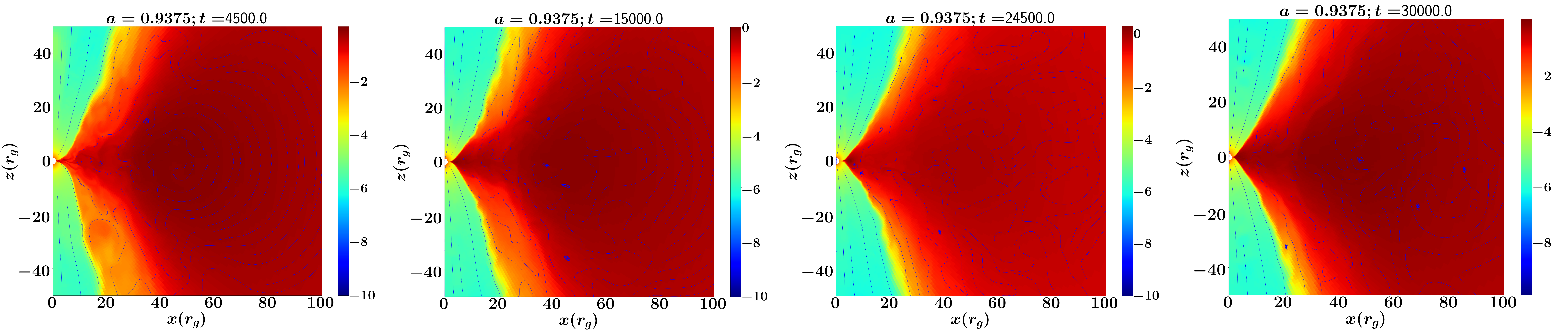}\label{sane}
}
\caption{Logarithmic density contours at various time for (a) SANE and (b) MAD simulations with magnetic field streamlines, constructed using the GRMHD code \texttt{BHAC}. Initially accretion proceeds similarly in both cases. However, around time $t=4900r_g/c$, where $r_g=GM/c^2$ is the gravitational radius with $M$ being the mass of BH, $G$ the Newton's gravitation constant, and $c$ the speed of light, we see the first flux eruption in MAD, in which matter is kicked away from the BH. These eruptions then repeat throughout the evolution of the accretion flow. Accumulation of strong magnetic fields is evident near the BH for MAD, leading to the formation of magnetic barrier which cuts off the accretion flow to the BH.}
     \label{cont}
\end{figure*}

\begin{figure}[ht!]
\centering
\includegraphics[width=0.49\textwidth]{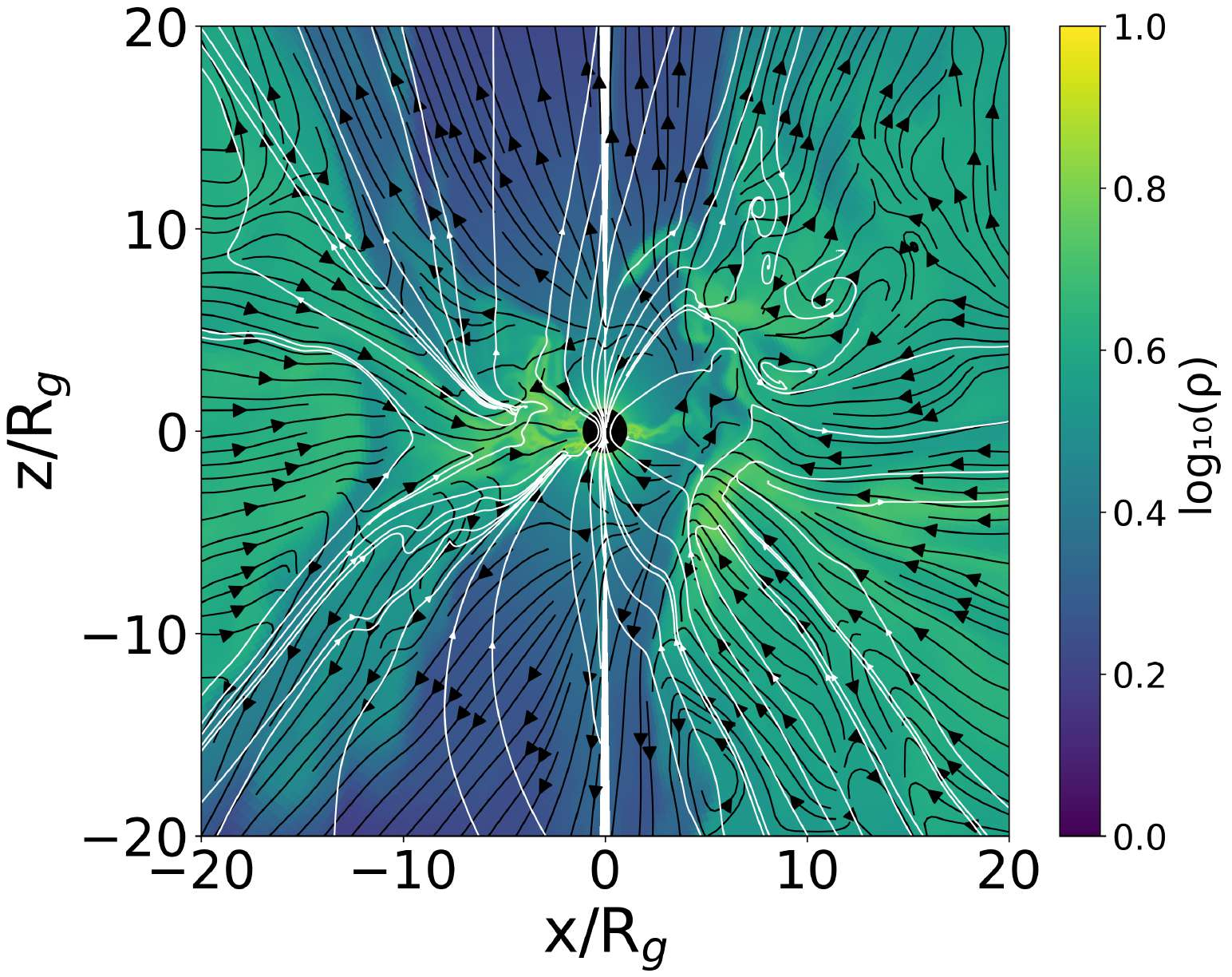}
\hfill
\includegraphics[width=0.49\textwidth]{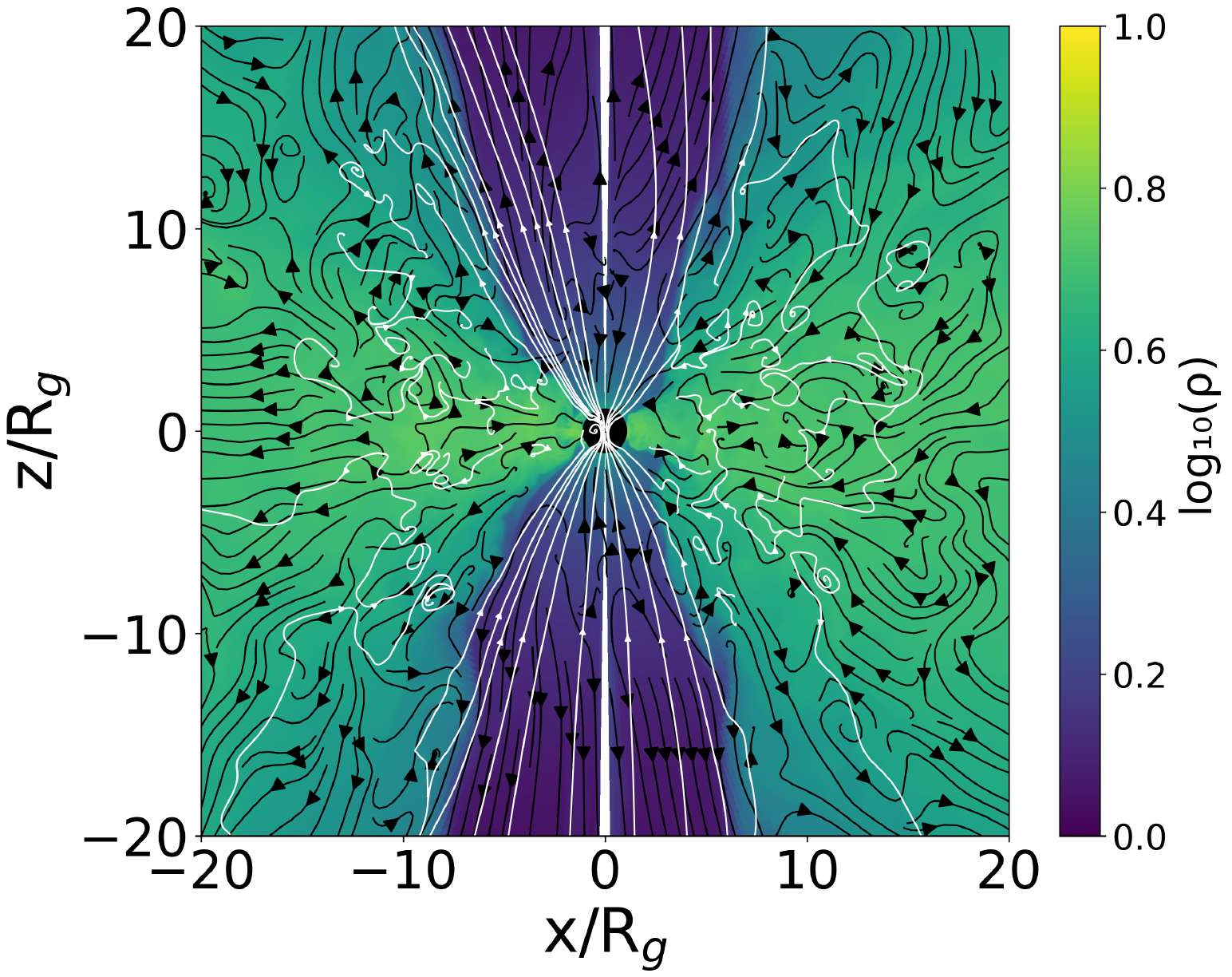}
\caption{Instantaneous density color contours of 3D simulations in x-z plane for (left) MAD at $t=23690 r_g/c$ and (right) SANE at $t=22620 r_g/c$. The color shows density values in logarithmic scale, the white lines are the magnetic field lines and the black arrows indicate velocity directions. [Online animation: See our high resolution movie of density evolution with time on YouTube playlist: \url{https://www.youtube.com/watch?v=SCg3zeG5iUI&list=PLHH19kmpHHtG6ONUUhqppc5gKlESPKLG-}]}
\label{fig:density_contours_3D}
\end{figure}

Magnetic fields are initiated in the simulation domain by using magnetic vector potentials of various forms, each resulting in different steady state simulation configurations. Two of the most commonly used vector potentials are \citep{rn2022}:
\begin{enumerate}
    \item SANE: $A_\phi=\max[\rho/\rho_0-0.2,0]$,
    \item MAD: $A_\phi=\max[(\rho/\rho_0)\exp(-r/400)(r/r_{in})^3\sin^3\theta-0.2,0]$.
\end{enumerate}
As evident, the initial magnetic field configuration due to the above vector potentials is poloidal. The magnetic field evolves very rapidly in MAD simulations, leading to fast build-up of magnetic flux near the BH. This accumulated flux leads to higher magnetic pressure on the incoming matter, eventually blocking the accretion flow in that direction effectively \citep{raha2023, raha2025}. Subsequent flux build-up leads to flux-eruption events in which matter (and field, due to flux-freezing) is kicked outwards (Figs.~\ref{mad}, ~\ref{fig:density_contours_3D} left panel). This leads to the reduction of magnetic flux allowing the accretion to proceed \citep{mp-bm}. This cycle is the characteristic feature of MAD simulations and is not observed in SANE simulations (as evident in Figs~\ref{sane} and ~\ref{fig:density_contours_3D} right panel), due to paucity of magnetic flux near the BH for the latter. The higher variability in accretion rate and magnetic flux in MAD compared to SANE (see Table ~\ref{tab:parameters}) also supports this feature.

\begin{table*}[h!]
\caption{Simulated properties of different 3D models for $a=0.998$}
\label{tab:parameters}
\small % Even smaller font
\setlength{\tabcolsep}{1.5pt} % Reduce spacing between columns
\centering
\begin{tabular}{lcccccccccc}
\hline\hline
State & $|<v^r>_\rho|$ & $\Gamma_{max}$ & $<B_{tot}>_\rho$ & $<\lambda>_\rho$ & $<\Omega/\Omega_K>_\rho$ & $<\sigma_m>_\rho$ & $<$plasma-$\beta>_\rho$ & $<h/r>_\rho$ & $\dot{M}_{r_H}$ & $\phi_{r_H}$ \\
 & (at $r_H$) & & (at $r_H$) & (at $r_H$) & & (at $r_H$) & (at $r_H$) & & variability & variability \\
\hline
MAD & 0.10 & 3.23 & 1.06 & 0.60 & 0.55 & 5.89 & 0.12 & 0.2-0.6 & 0.236 & 0.150 \\
SANE & 0.03 & 2.57 & 0.42 & 1.77 & 0.95 & 0.13 & 3.24 & 0.2-0.3 & 0.180 & 0.077 \\
\hline
\end{tabular}
\small
\begin{justify}
%\justifying 
\textit{Note:} All quantities in this table are derived from 3D GRMHD simulations (time-averaged over 20,000-25,000$r_g/c$):
   $|<v^r>_\rho|$: Radial velocity at the horizon in units of c;
   $\Gamma_{max}$: Maximum Lorentz factor of accelerated jets;
   $<B_{tot}>_\rho$: Total magnetic field strength at the horizon in code units;
   $<\lambda>_\rho$: Specific angular momentum at the horizon;
   $<\Omega/\Omega_K>_\rho$: Ratio of angular velocity to Keplerian angular velocity at outer radius;
   $<\sigma_m>_\rho$: Magnetization parameter at the horizon;
   $<$plasma-$\beta>_\rho$: Ratio of gas pressure to magnetic pressure;
   $\dot{M}_{r_H}$ variability: Mass accretion rate variability (computed as $\sigma_{\dot{M}}/\mu_{\dot{M}}$);
   $\phi_{r_H}$ variability: Magnetic flux variability (computed as $\sigma_\phi/\mu_\phi$).
All quantities are shell-averaged over $\theta$ and $\phi$ directions.
\end{justify}

\end{table*}

The differences in various properties of the two accretion states for a highly spinning BH ($a=0.998$) are described in detail in \citealt{raha2025} (see Table \ref{tab:parameters}).

\section{Black hole unification across mass scales}

It has been argued that the underlying physical ingredients required to power jet formation and the underlying accretion flow in BHs of different masses is the same. This leads to quantities like X-ray ($L_X$) and radio ($L_R$) luminosities being scaled with the mass of the central BH ($L_X\propto L_R^\alpha M^\beta$, where $M$ is the mass of the central gravitational object and $\alpha$ and $\beta$ are the power law exponents) \citep{sera}.

Utilizing the VLBI capabilities of SKA, the outflow power of radio jets and their morphologies from stellar and supermassive BHs can be estimated to a high accuracy. The full commensal wide and deep surveys of SKA1-MID band 3 and band 5 \citep{agudo} will have highly sensitive polarimetric capabilities which will enable power estimations of AGNs and distinguish between jet and star formation emission over a wide range of red-shifts. 

Stellar mass BHs are found mainly in the low-hard and high-soft state. The low-hard state is characterized by a persistent radio jet, while the high-soft state has strong thermal X-ray emission from the accretion disk and the radio emission is relatively weaker. Thus, accreting stellar mass BHs in hard state \citep{raha2024} form ideal candidates for SKA-VLBI to estimate their outflow power and compare it with that of AGN based simulations.
Replicating the scaling power-law between X-ray and radio luminosities ($\alpha$ and $\beta$) obtained from the above observations with GRMHD simulations using appropriate parameters will further improve the understanding of the underlying accretion flow dynamics in these systems. 

\section{Magnetic field topology}

\begin{figure*}
\centering
\begin{minipage}[t]{0.45\textwidth}
\vspace{-2cm}
\scriptsize
\begin{tabular}{c|cccc}
\hline\hline
& \multicolumn{2}{c}{MAD} & \multicolumn{2}{c}{SANE}  \\
\cline{2-3} \cline{4-5}
& BHAC~(2D) & H-AMR~(3D) & BHAC~(2D) & H-AMR~(3D) \\
\hline
$a$ & $0.9375$ & $0.998$ & $0.9375$ & $0.998$ \\
\hline
$\eta$ & 3-9 & 1-10 & 2.3-4 &0.7-1   \\
\hline
\end{tabular}
\captionof{table}{Comparison of outflow efficiencies for different BH spins and magnetic field configurations around the event horizon.}
\label{tab:efficiency}
\end{minipage}
\hfill
\begin{minipage}[t]{0.45\textwidth}
\centering
\includegraphics[width=\textwidth]{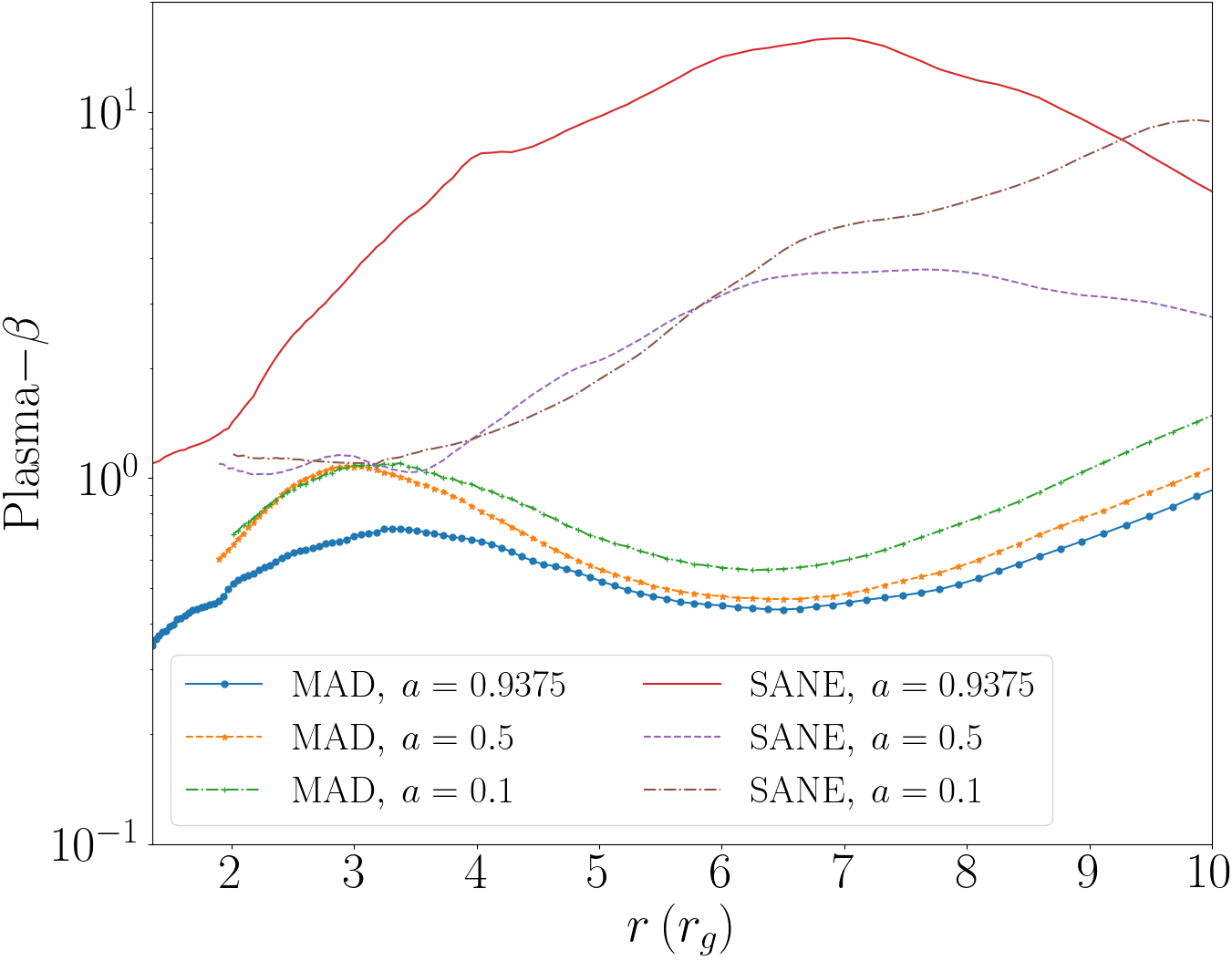}
\caption{Plasma-$\beta$ profiles for different BH spins for MAD and SANE, calculated from simulations run using \texttt{BHAC}. Here $r$ is measured in units of $r_g$.}
\label{beta}
\end{minipage}
\end{figure*}

The jet power according to the BZ mechanism is given by \citep{meier01},
\begin{equation}
    P=\frac{\Phi^2\Omega^2}{32c},
\end{equation}
where $\Phi$ is the (poloidal) magnetic flux  and $\Omega$ is the angular velocity of the jet launching region.
As explained in Sec.~\ref{grmhd}, since MAD simulations have higher $\Phi$, the outflow power is also higher for a given BH spin as compared to SANE simulations. Thus, given the outflow luminosity it can be determined whether the system is MAD or SANE. To calculate the outflow efficiency from simulations, the following quantities are utilised \citep{fluxerr}:
\begin{enumerate}
    \item Accretion rate: $\Dot{M}=-\int\sqrt{-g}\rho u^r\mathrm{d}\theta \mathrm{d}\phi$,
    \item Efficiency: $\eta=P_c/\dot{M}$, where $P_c=\Dot{M}-\Dot{E}$ is the outflow power and $\Dot{E}=\int \sqrt{-g}T^r_t\mathrm{d}\theta \mathrm{d}\phi$, where $T^{\mu}_{\nu}=(\rho+p+u_g+b^2)u^{\mu}u_{\nu}+(p+b^2/2)\delta^\mu_\nu-b^{\mu}b_{\nu}$ is the stress-energy tensor.
\end{enumerate}
Here, $g$ is the determinant of the background metric, $\rho$ is the disk density, $p$ is the pressure of the flow, 
$\gamma$ is the adiabatic constant, $u_g=p/(\gamma+1)$ is the internal energy of the fluid, $u^{\mu}$ and $b^{\mu}$ are the four-velocity and four-magnetic field respectively, and $b^2=b^{\mu}b_{\mu}$. As evident in Table~\ref{tab:efficiency}, MAD systems have more outflow efficiency than SANE systems. The variation in the efficiency of the flow is due to the underlying time evolution of the simulation domain.

However, another important study to verify these efficiencies will be to determine the magnetic field characteristics of the system. MAD systems are magnetically dominated, with their plasma-$\beta$ (gas pressure/magnetic pressure) <1, while for SANE, plasma-$\beta$ >1 (Fig.~\ref{beta}, Table ~\ref{tab:parameters}) \citep{mp-bm, raha2025}. 
MAD systems are also dominated by structured poloidal magnetic fields (Figs~\ref{mad} and ~\ref{fig:density_contours_3D} left panel) which facilitate jet launching. 
Thus, luminosity measurements and magnetic field topology in jets inferred from observations, combined with the knowledge of underlying BH mass and spin will conclusively determine whether a system has MAD or SANE magnetic field characteristics.

Polarization measurement studies like linear polarization (LP) (Stokes Q and U parameters), circular polarization (CP) (Stokes V parameter) and rotation measure (RM) will reveal several fundamental characteristics of the radio jet. For example, the low energy cutoff of the electron energy distribution is the ratio of CP and RM. This parameter is crucial for the study of synthetic spectra due to non-thermal electrons. The total LP defined by $L=\sqrt{U^2+Q^2}$, can be very useful in quantifying the magnetic field strength and, thus, the regions with plasma-$\beta$<1, as these regions will have high $L$ values. In addition to this, polarization position angle (PPA) given by $\theta=(1/2)\tan^{-1}(U/Q)$ can help in estimating the magnetic field geometry. The polarization fraction ($\tilde{p}=\sqrt{L^2+V^2}/I$, where $I$ is the total polarization) can be used to distinguish between regions of plasma-$\beta$<1 and >1, as in the latter, the system is gas pressure dominated, leading to very low values of $\tilde{p}$.

CP analysis in particular will help in calculating the magnetic field strength, and the vector-ordering of the field. This will help in determining the magnetic flux carried by the jet, which can be directly compared with simulation results of $\Phi$.
Although CP studies are generally difficult due to its weak and highly variable nature, the large instantaneous wavelengths and high sensitivity of SKA are required for its accurate measurement, as long as the polarization purity ($\sim 0.1\%$ of LP and $\sim 0.01\%$ of CP) \citep{agudo} is maintained.
Polarization studies carried out using SKA will thus result in in-depth and comprehensive knowledge of the dynamics of the radio jet and can reveal information about its generation mechanisms as well.

However, the polarization of radiation originated in the jet region can be altered due to Faraday rotation by the magnetic fields confining/collimating the jet. Nevertheless, given the magnetic field strength and geometry, radiative GRMHD simulations can be used to estimate the RM of the jet region \citep{rm-2017}, by appropriate modeling of the electron number density. The observed polarization can then be corrected by  removing the effect of the Faraday rotation, using the RM obtained from the GRMHD simulations. This will lead to a clearer picture of the radiation emitting regions in the interior of the jet.

\section{Jet collimation due to feedback from ambient medium interaction}

Interaction of jets from AGNs with their ambient medium has been shown to affect star formation, outflow of matter, spectral features etc. in galaxies \citep{agn-jet}. The ambient medium and feedback from the jet can also alter the jet structure and lead to collimation.

Collimation of jets can be explained by magnetic tension force due to toroidal magnetic fields close to the BH (above the Alfv\'en radius). But certain systems, like 3C 84, show collimation of jets into cylinders at larger distances as well \citep{gio}. It has been shown using GRMHD simulations that jets undergo significant collimation due to interactions with disk winds and the ambient medium at large distances from the BH. The pressure of the ambient medium and the radiative cooling present in the system play a key role in the extent of collimation at a given distance from the BH \citep{cyl23}. 

The high resolution and sensitivities of the order of $\mu$~Jy/beam of SKA will be instrumental in determining the extent of collimation at a given distance from the BH. As jets are mainly collimated by toroidal magnetic fields, a measure of toroidal and poloidal magnetic fields in the jet region can help in our understanding of this jet collimation mechanism. The PPA can be used to estimate the strength of toroidal magnetic field along the jet axis. PPA for toroidal field dominated regions will deviate substantially from regions with poloidal fields. The enhanced polarization capabilities of SKA will thus enable us in distinguishing toroidal and poloidal magnetic field dominated regions and, thus, aid in studying jet collimation. The observed jet collimation can be compared with the collimation obtained from non-radiaitve GRMHD simulations. If the two collimations do not corroborate, it shows the presence of significant radiative cooling in the system, leading to further collimation. The extent of this radiative cooling driven collimation can then be inferred by comparing jet characteristics obtained from radiative GRMHD simulations by modeling the observed system appropriately.
This information, combined with the characteristics of the ambient medium which can be effectively probed by SKA1-LOW, SKA1-MID and SKA1-SUR (bands 1–3) \citep{unravel}, will reveal important features about the system when modeled with GRMHD simulations about the inner cooling mechanisms, magnetic fields and central engine characteristics. 

AGN jets typically feature knots and their dynamics has been used in studying the interaction of the jet with the ambient medium using spectral index maps \citep{park-24}. The superior spectral resolution and sensitivity of SKA will allow probing of jet region closer to the central BH, thus leading to more data about the jet-ambient medium interaction. This will allow for better modeling of the BH system using GRMHD simulations and consequently better estimates of the intrinsic properties of the underlying accretion flow.

\section{Multi-wavelength analysis and future prospects}

The radio emission from BH jets is often accompanied with emission for the accretion disk in the optical-UV-X-ray range. Correlated studies of radio observations from SKA and other telescopes in the appropriate frequency range will reveal important characteristics of the accretion flow as well. This can then be modeled based on GRMHD simulations. For instance, as the disk-jet system is symbiotic \citep{falcke-95,debbijoy-10,ULX20}, information of jet characteristics is correlated with complimentary accretion disk data. The radio and X-ray luminosities ($L_R$ and $L_X$ respectively) of BH X-ray binaries are indeed shown to have correlation ($L_R\propto L_X^b$ with $b\sim0.7$) \citep{fender}. With the enhanced resolution and sensitivities of SKA, the $L_R$ of various BH accretion sources can be estimated with great accuracy. The X-ray observations of these sources can possibly extend the aforementioned correlation to more sources and higher red-shifts. 

Correlation of $\gamma$-ray and radio luminosities for blazars has been shown to exist for several sources. \cite{wu-2014} showed the correlation between $L_R$ at 5 GHz and $\gamma$-ray luminosities using Fermi/LAT. Such studies help in identifying the beaming characteristics of the blazars and in the debeaming of the observed beamed luminosities to reveal the intrinsic outflow power of the system \citep{debbijoy-10,nemmen-12,BL19}. Correlating blazar observations using SKA with Fermi/LAT can lead to identification of the beaming properties of blazar sources at various red-shifts due to the high sensitivity of SKA.

Future SKA enhancements may also help in understanding differences between dynamical characteristics of accretion disks. For example, magnetized accretion flows feature magnetic fields of varying magnitudes. Depending on the magnetic field structure, distance from the BH and its spin, the difference in properties like outflow power, plasma-$\beta$ can be quite small (Fig. \ref{beta}, Table \ref{tab:efficiency}). Therefore, to distinguish such sources, higher spectral resolution and polarization calibration is required. This will help in identifying small changes/differences in magnetic field characteristics. Also the synchrotron emission is dependent on the magnetic field strength of the flow. To cover a wide range of synchrotron luminosity (and thus magnetic field strength), higher frequency coverage is also important. The peak of synchrotron emission is also sensitive to the mass of the central BH. Thus a higher frequency coverage will also help in targeting lighter/heavier BH systems.

\section{Conclusion}

The high resolution and sensitivity of SKA will open new avenues for accretion disk-jet systems. With its deep field survey capabilities, SKA will be able to survey high redshift ($z\sim10$) galaxies as well, thus increasing our sample space for studying the properties of accretion disk-jet systems around BHs. These observations will enable better understanding of BH unification ideas across mass scales. The corresponding polarization measures with high purity ($0.01\%-0.1\%$) will help in constraining the magnetic field structures. Observation of jet structure and morphologies at different length scales will further help in exploring jet dynamics, its interaction with the ambient medium and jet launching mechanisms. Modeling and comparison of these systems with GRMHD simulations will reveal important information about various system characteristics like radiative cooling processes, mass and spin of the BH, detailed magnetic field and jet topology close to the BH etc.

\section*{Acknowledgments}

We thank Avishek Basu (Jodrell Bank Centre for Astrophysics, Manchester) for discussions regarding the observational signals of various magnetic field characteristics within SKA capabilities. MP and RR acknowledge the Prime Minister’s Research Fellows (PMRF) scheme for providing fellowship.

\bibliographystyle{abbrvnat-maxbibnames4}

\bibliography{chapter}

\end{document}